\documentclass[reprint,amsmath,amssymb,aps,pra,nofootinbib,longbibliography]{revtex4-1}

\usepackage{bm}
\newcommand{\kF}{\ensuremath{k_\mathrm{F}}}
\newcommand{\dd}{\mathrm{d}}

\usepackage{graphicx}

\usepackage[hidelinks]{hyperref}
\usepackage{siunitx}

\begin{document}

% Info
\title{Transition from supersonic to subsonic waves in superfluid Fermi gases}

\author{Senne Van Loon}
\email{Senne.VanLoon@UAntwerpen.be}
\author{Wout Van Alphen}
\author{Jacques Tempere}
\author{Hadrien Kurkjian}
 \affiliation{TQC, Universiteit Antwerpen, Universiteitsplein 1, B-2610 Antwerpen, Belgi\"e}

\date{\today}

\begin{abstract}
	We study the propagation of dispersive waves in superfluid Fermi gases in
	the BEC-BCS crossover. Unlike in other superfluid systems, where dispersive
	waves have already been studied and observed, Fermi gases can exhibit a
	subsonic dispersion relation for which the dispersive wave pattern appears
	at the tail of the wave front. We show that this property can be used to
	distinguish between a subsonic and a supersonic dispersion relation at
	unitarity.
\end{abstract}

\maketitle

\section{Introduction}

Cold atomic gases have given a new boost to the research on superfluids. Using
the high level of experimental control offered by these systems, the propagation
of first \cite{Ketterle1997,hoinka2017goldstone} and second sound
\cite{sidorenkov2013second} has been observed, the superfluid fraction has been
measured \cite{sidorenkov2013second}, the dissipationless flow of an impurity
below the critical velocity was demonstrated \cite{Salomon2015}, and the damping
of phonons has been precisely measured in Bose gases
\cite{Dalibard2002,Davidson2002} and clearly related to elementary three phonons
processes \cite{LandauKhal1949,beliaev1958st}.

In this context, cold gases of paired fermions have attracted special attention
due to the possibility of tuning the interaction strength using a Feshbach
resonance \cite{Grimm2003}. This degree of freedom allowed for the observation,
unique among superfluid systems, of a resonantly interacting gas in the
so-called unitary limit \cite{Thomas2002}. A specificity offered by the
controllable interactions is that the sound branch changes from a supersonic
dispersion relation in the Bose-Einstein condensate (BEC) limit, where the pairs
are tightly bound dimers, to a subsonic one in the Bardeen-Cooper-Schrieffer
(BCS) limit of weakly correlated pairs \cite{kurkjian2016concavity}. Cold Fermi
gases are then one of the rare homogeneous superfluid systems in which a
subsonic dispersion relation can be observed (others being helium at high
pressure \cite{rugar1984accurate} and a spin-orbit coupled BEC
\cite{lin2011spin}). Since dissipative effects are weak in low temperature
superfluid Fermi gases \cite{kurkjian2016landau,Kurkjian2017PRL}, waves
propagate much longer than in a viscous medium. The long time behavior of a wave
packet is then governed by dispersive effects
\cite{Whitham1974,el2016dispersive}, and a specific behavior, never before
observed in a superfluid, is expected for a subsonic dispersion
\cite{sprenger2017shock}.

Describing this dispersive hydrodynamics in a Fermi gas is a nontrivial task.
Since high-amplitude waves excite the pair internal degrees of freedom, there
exists no simple equivalent of the bosonic Gross-Pitaevskii equation able to
describe the nonlinear wave dynamics and relate it to well-studied mathematical
models such as the Kortweg-de Vries equation \cite{KdV,shearer2017}. Here we
study wave propagation in two limiting cases where rigorous wave equations can
be derived from first principles.

In Sec.~\ref{sec:Dispersion}, we study small amplitude waves completely
characterized by the dispersive spectrum. Due to dispersion, the plain wave
front that would propagate after a perturbation in a nondispersive medium is
perturbed by the formation of an oscillatory train. The position of these
oscillations with respect to the wave front depends on whether the bending of
the sound branch is supersonic or subsonic, and thus changes when the
interactions are tuned from the BEC to the BCS regime.

In Sec.~\ref{sec:shockwaves}, we study the propagation of large-amplitude
long-wavelength perturbations using the nonlinear wave equation derived in
Ref.~\cite{Klimin2015}. With an initial perturbation in the form of a density
depletion, we show the appearance of a narrow solitary edge traveling slower
than the speed of sound behind the wave front. The secondary peaks caused by
dispersion are smoothened by nonlinear effects but remain visible. This behavior
is reminiscent of the dispersive shock waves observed in Bose gases
\cite{hoefer2006dispersive,Hoefer2008DSW}.

Finally, we show how these phenomena can be used to settle the ongoing debate on
the interaction regime at which the collective branch changes from supersonic to
subsonic. Calculations in the random-phase approximation (RPA)
\cite{diener2008quantum,kurkjian2016concavity},  in an effective Lagrangian
approach \cite{Manuel2009}, in a $4-\epsilon$ expansion
\cite{rupak2009density}, or Monte-Carlo simulations \cite{salasnich2008extended}
predict a supersonic branch at unitarity, while a density functional method
\cite{zou2017low} finds it subsonic. To date, there is no measurement that can
settle this controversy, although the supersonic or subsonic nature of the sound
branch controls several important macroscopic properties of the gas, in
particular its dissipative properties \cite{LandauKhal1949}. Here we show that
dispersive waves can be used to obtain such a measurement using state-of-the-art
experimental techniques to create small-sized perturbations \cite{Roati2017} and
to perform high-resolution imaging \cite{bloch2012quantum}.

\section{Linear dispersive waves \label{sec:Dispersion}}

\begin{figure*}
	\centering
	\includegraphics[scale=1]{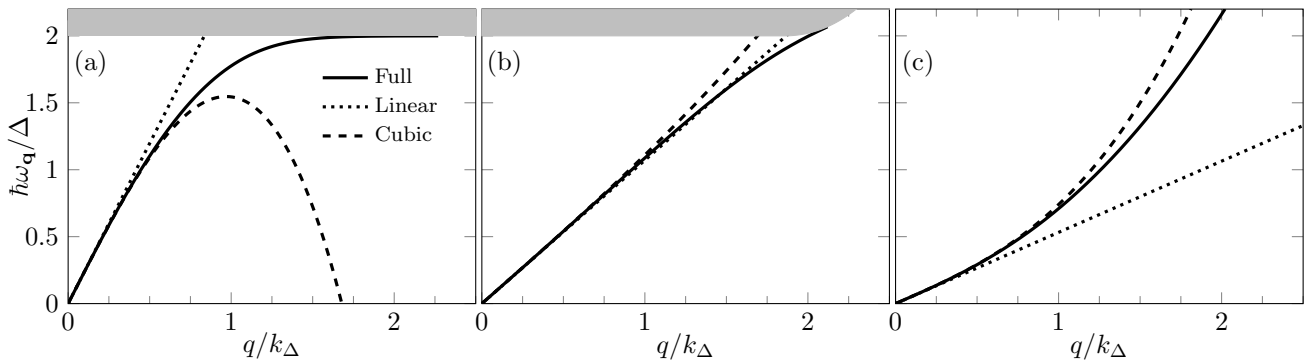}
	\caption{ \label{fig:Dispersion}
	The RPA dispersion relation of the collective excitations is plotted (a) in
	the BCS regime ($1/\kF a=-1$), (b) at unitarity ($1/\kF |a|=0$), and (c) in
	the BEC regime ($1/\kF a=1$). The full line corresponds to the full numeric
	solution of the dispersion, which is compared to its linear (dotted) and
	cubic (dashed) approximation at low $q$. The gray area shows the
	pair-breaking continuum. Units of the superfluid order parameter $\Delta$
	and $k_\Delta = \sqrt{2m\Delta}/\hbar$ are used respectively for the energy
	and the wavenumber.}
\end{figure*}

At low momentum, the dispersion relation of the sound branch of a superfluid can
be written generically as
\begin{equation}
\hbar \omega_{\bf{q}} = 
\hbar c q \left[1 
+ \frac{\gamma}{8} \left(\frac{\hbar q}{m c} \right)^2
+ O\left(\frac{\hbar q}{m c}\right)^4  \right].
\label{eq:omlowq}
\end{equation} 
In this expression, $c$ is the speed of sound, found from the density $\rho$ and
the chemical potential $\mu$ of the gas by the hydrodynamic relation $m c^2=\rho
\dd\mu/\dd\rho$, $m$ is the mass of the particles, and $\gamma$ is a
dimensionless parameter controlling the cubic correction to the linear spectrum.
In a superfluid Fermi gas, the speed of sound is known experimentally for any
interaction strength from the measurements of the equation of state
$\mu=\mu(\rho,a)$, with $a$ the $s$-wave scattering length
\cite{Salomon2010,Zwierlein2012}. For the coefficient $\gamma$, which depends on
the microscopic physics of the system and trapping geometry, there are however
only theoretical predictions. For homogeneous gases, several predictions of a
positive \cite{diener2008quantum,kurkjian2016concavity,Manuel2009,
rupak2009density,salasnich2008extended} or negative \cite{zou2017low} $\gamma$
coexist at unitarity ($|a|=\infty$) but only the RPA prediction of $\gamma$
exists in the whole BEC to BCS crossover \cite{kurkjian2016concavity}. In
particular, the RPA finds $\gamma$ to be negative for $1/k_{\rm F}a<-0.14$ and
positive above. At higher momentum, the full dispersion relation
$q\mapsto\omega_q$ was again only predicted within the RPA; it is obtained by
numerically solving the RPA implicit equation
\cite{combescot2006,diener2008quantum} (see Eq.~(1) in
Ref.~\cite{kurkjian2016concavity}). This dispersion relation is visualized in
Fig.~\ref{fig:Dispersion} for different interaction regimes.

In this work we explain how dispersive waves can be used to measure the
coefficient $\gamma$. Our starting point is the Schr\"odinger equation that
governs the propagation of a plane wave of momentum $\bf{q}$:
\begin{equation}
\left( \mathrm{i} \partial_t - \omega_{\bf{q}} \right) \psi =0,
\label{eq:WaveEq}
\end{equation}
where $\psi \in \mathbb{R}$ represents a perturbation of the superfluid density
$\rho=\rho_0(1+\psi)^2$. This very intuitive equation is in fact rigorously
demonstrated for a superfluid Fermi gas by writing down, in a functional
integral formalism, a quadratic Lagrangian for the phase and amplitude of the
superfluid order parameter, as is done explicitly in Appendix~\ref{AppEOM}.
Replacing $\omega_{\bf{q}}$ by its cubic approximation \eqref{eq:omlowq} and
restricting to one dimensional right-propagating waves Eq.~\eqref{eq:WaveEq}
takes the form
\begin{equation}
\partial_t^{} \psi = -c \partial_x^{} \psi + \frac{\gamma \hbar^2}{8m^2 c} 
\partial_x^3 \psi,
\label{eq:KdV}
\end{equation}
which is nothing else than a linearized Kortweg-de Vries equation
\cite{boussinesq1877essai,KdV}. The propagation of unidimensional waves in
(quasi)homogeneous space can be studied in box potentials \cite{Zwierlein2017},
provided the transverse size of the box is much larger than the wavelength of
the perturbation \cite{Hadzibabic}. In elongated harmonic traps, the dispersion
of phonons is expected to be concave in the BEC limit, as in a weakly
interacting Bose gas \cite{Fedichev2001}, so that no transition from subsonic to
supersonic waves should occur.

We study the propagation of an initial Gaussian perturbation of the superfluid
density
\begin{equation}
\psi(x,t=0) = \zeta \, \mathrm{e}^{-\frac{x^2}{2 \sigma^2}},
\end{equation}
where the amplitude $\zeta$ is chosen small enough for the linear differential
equation \eqref{eq:KdV} to remain valid. Upon rescaling the distances to the
width of the perturbation $\sigma$ and the times to its duration $\sigma/c$,
there remains a unique parameter describing the propagation of waves under
Eq.~\eqref{eq:KdV}, namely the coefficient of the third order derivative
$\gamma\hbar^2/8m^2c^2\sigma^2$. This parameter thus controls the time after
which the dispersive effects become important 
\begin{equation}
	t_\mathrm{sep} = \frac{\sigma}{|c(\tilde{q})-c|}
		 = \frac{2\sigma}{c |\gamma|} \left( \frac{m c \sigma}{\hbar} \right)^2.
	\label{eq:tsep}
\end{equation}
Here $c(q) = \omega_{q}/q$ is the phase velocity of the waves with momentum $q$
and $\tilde{q}=2/\sigma$ is the typical wavenumber of the high-momentum waves in
the perturbation. At time $t=t_{\mathrm{sep}}$ the waves with wavenumber
$\tilde{q}$ have traveled away from the main wave front across a distance
$\sigma$, leading to the formation of an oscillatory train. The width $\sigma$
should be chosen small enough for the separation time to remain within
experimental reach, yet large enough for the cubic expansion \eqref{eq:omlowq}
to be valid. 

\begin{figure}
	\centering
	\includegraphics[scale=1]{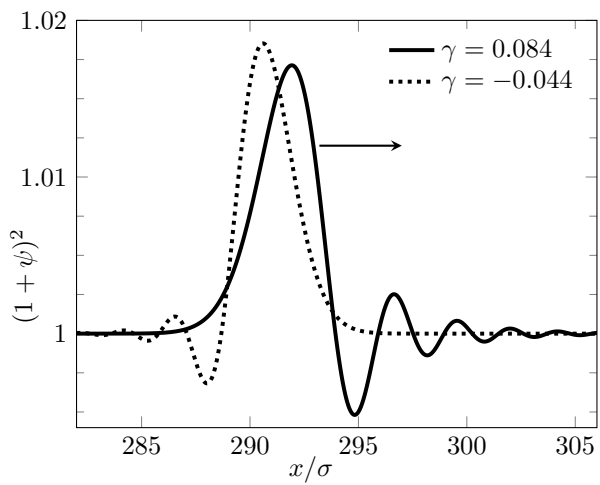}
	\caption{ \label{fig:WaveUni}
	A comparison of dispersive waves for different predictions of the cubic
	coefficient $\gamma$ at unitarity ($1/\kF a= 0$). Both functions are
	solutions to Eq.~\eqref{eq:KdV}, starting from a Gaussian perturbation with
	$\zeta=0.02$ and $\sigma = 2.5 \hbar/mc$. For the solid curve the analytic
	RPA prediction $\gamma = 0.084$ is used, while the dotted line is drawn for
	$\gamma = -0.044$, predicted by Zou \textit{et al.} \cite{zou2017low}. The
	dispersive waves are shown at a time $t=t_\mathrm{sep}$ and we omitted the
	symmetric left-traveling wave for visibility.}
\end{figure}

In Fig.~\ref{fig:WaveUni} we show the dispersive waves at unitarity ($1/\kF a=
0$) for $\sigma = 2.5 \hbar/mc$, comparing the prediction of $\gamma$ of
Ref.~\cite{zou2017low} to the expression of the RPA.
The difference between supersonic and subsonic dispersive waves is clearly
visible. For the positive $\gamma$ predicted by the RPA, secondary oscillations
appear at the leading edge of the traveling wave, while for a negative $\gamma$
they appear at the trailing edge. Observing the location of these secondary
oscillations is thus enough to predict the sign of the cubic term in the
dispersion.

\begin{figure}
	\centering
	\includegraphics[scale=1]{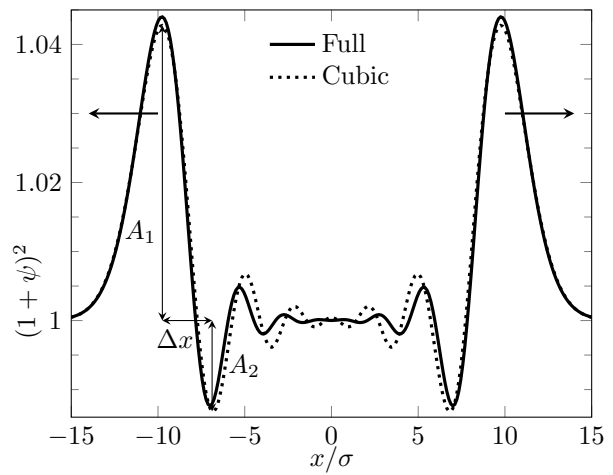}
	\caption{ \label{fig:WaveBCS}
		Dispersive waves in the BCS regime ($1/\kF a= -1$) at a time $t=2
		t_\mathrm{sep}$, starting from an initial Gaussian perturbation with
		$\zeta=0.05$ and $\sigma=3.2 \hbar/mc$. The cubic approximation of the
		dispersion (dotted line) is compared to the full numeric solution (solid
		line). The cubic approximation becomes worse moving away from the
		primary wave front, where high-momentum waves dominate.}
\end{figure}

In the BCS regime $\gamma$ is certainly negative, offering a system with a
subsonic dispersion. This can be seen in Fig.~\ref{fig:WaveBCS}, where secondary
oscillations appear behind the traveling wave front. 
There we compare dispersive waves generated by the full dispersion relation to
those generated by its cubic approximation. Both solutions coincide close to the
primary wave front, but start to differ further away, where higher wavenumbers
become important and the cubic expansion is not valid anymore.

\section{{Shock waves}}
\label{sec:shockwaves}

To go beyond the small amplitude approximation and account for nonlinear effects
in our physical situation, we now search for a nonlinear wave equation.
Obtaining such an equation in a strongly interacting superfluid, and especially
in a superfluid of fermions, is a difficult task. First, the (nonlinear)
Korteweg-de Vries equation (or its extensions that include an arbitrary
amplitude dependence of the speed of sound \cite{shearer2017}), which would seem
like a natural generalization  of Eq.~\eqref{eq:WaveEq}, describes separately
right- and left-travelling waves \cite{salasnich2011supersonic} such that it
does not describe our situation where a perturbation initially at rest splits
into two counter-propagating waves and where important nonlinear effects take
place during the separation stage. To correctly describe counter-propagating
waves, we need at least a system of two coupled nonlinear equations, as for
example the (complex) Gross-Pitaevskii equation.

Second, deriving a fermionic equivalent of the Gross-Pitaevskii equation is
arduous because high-amplitude excitations excite the internal degrees of
freedom of the fermion pairs. A first possibility is to use Bogoliubov-de Gennes
equations of motion, which are a large set of coupled nonlinear equations
\cite{deGennes1966}. Alternatively, Ref.~\cite{Klimin2015} achieves it by
restricting to long-wavelength and low-energy perturbations. The ensuing
nonlinear wave equation on the superfluid order parameter $\Psi$ takes the
following form
\begin{equation}
\begin{aligned}
\mathrm{i} D(\vert \Psi \vert^2) \, \partial_t \Psi = &- C \, \partial_x^2 \Psi
+ Q \, \partial_t^2 \Psi + A(\vert \Psi \vert^2) \Psi \\
& + \left(E \, \partial_x^2 \vert \Psi \vert^2 
- R \,  \partial_t^2 \vert \Psi \vert^2 \right) \Psi,
\label{eq:EFT}
\end{aligned}
\end{equation}
where the coefficients $C$, $E$, $Q$, $R$, and the functions $A$ and $D$ of the
wave intensity are given in Appendix~\ref{app:EFT} as integrals over the
fermionic degrees of freedom. Unlike the phenomenological system based on
hydrodynamics of Ref.~\cite{salasnich2011supersonic} this equation is derived
from first principles by resumming the infinite series of the slow fluctuations
of the order parameter, and naturally accounts for the fermionic contribution to
the wave dynamics. Unfortunately, since it truncates time and space derivatives
to second order, it does not describe correctly the dispersion coefficient
$\gamma$ in the BCS regime, where it depends on higher order derivatives.

\begin{figure}
	\centering
	\includegraphics[scale=1]{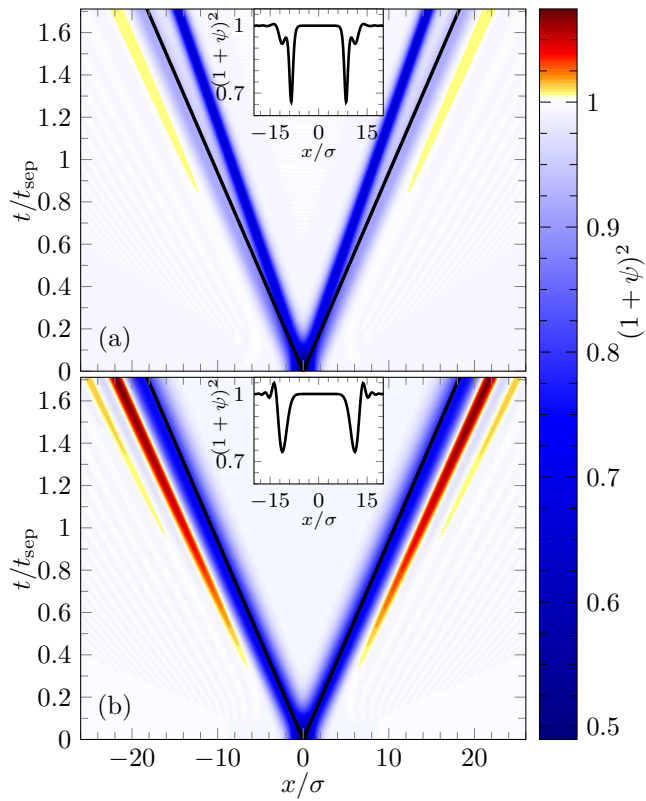}
	\caption{ \label{fig:cone}
	(Color online) The superfluid density following a localized initial
	perturbation in $x=0$ (with $\zeta=-0.3$ and $\sigma=1.14 \hbar/mc$) is
	shown in colors as a function of space (on the horizontal axis) and time (on
	the vertical axis), in the BEC regime ($1/\kF a= 2$). The black solid lines
	represent the light cone $x=\pm ct$. The top panel (a) shows the nonlinear
	evolution according to Eq.~\eqref{eq:EFT} while the bottom one (b) shows the
	linear dispersive evolution according to Eq.~\eqref{eq:WaveEq}.}
\end{figure}

In Fig.~\ref{fig:cone}, we use this equation to track the time evolution of a
large decrease ($\zeta=-0.3$) of the superfluid density in the BEC regime, and
compare it to the linear dispersive scenario of Eq.~\eqref{eq:WaveEq}. The
secondary oscillations caused by the supersonic dispersion are still visible at
the leading edge of the wave, but their amplitude is reduced. In the trailing
edge a major nonlinear feature appears: a narrow solitary edge travelling at a
constant speed. Since we chose an initial perturbation that depletes the
density, this speed is here smaller than the speed of sound such that the
dispersive oscillations and the solitary edge are separated by the light cone
$x=\pm ct$.

The behavior observed here is reminiscent of the dispersive shock waves observed
in dissipationless nonlinear media
\cite{el2016dispersive,hoefer2006dispersive,Hoefer2008DSW}. This is remarkable
since our complex nonlinear equation \eqref{eq:EFT} is quite different from the
Kortweg-de Vries equation with which dispersive shock waves are usually
described.

\section{Experimental observability \label{sec:Experiments}}

We now discuss the observability of the secondary oscillations generated by the
dispersive wave propagation in real experimental conditions. 

\subsection{Propagation time \label{sec:Numerical}}

To be able to see dispersive waves, one should wait a time of order
$t_\mathrm{sep}$ (defined in Eq.~\eqref{eq:tsep}). At time $t=t_\mathrm{sep}$,
no matter the value\footnote{The solution of the wave equation \eqref{eq:KdV}
can be made independent of $\sigma$ by changing $x$ to $x^\prime=(x-c t)/\sigma$
and $t$ to $\sigma t^\prime/c = t/\tilde{\sigma}^2$, with $\tilde{\sigma}=m c
\sigma/\hbar$. Then the linearized Korteweg-de Vries equation \eqref{eq:KdV}
takes the simple form $\partial_{t^\prime} \psi = \pm \partial_{x^\prime}^3
\psi$, where the sign is that of $\gamma$ and with the initial condition
$\psi(x^\prime,t^\prime=0) = \zeta \mathrm{e}^{-{x^\prime}^2/2}$.} of $m c
\sigma/\hbar$, the primary density excitation has an amplitude $A_1$ of about
95\% of the initial Gaussian perturbation $\zeta$, while the biggest secondary
peak $A_2$ is roughly 15\% of $\zeta$. At this time, the spatial distance
between the two peaks is approximately $\Delta x \simeq 2.6 \, \sigma$. 

To compare with experimental parameters, we use a system of $\mathrm{{}^6Li}$
atoms with a typical Fermi temperature $T_\mathrm{F}= 1\,\mu \mathrm{K}$. Using
a thin optical barrier to create the initial perturbation, one can reach a width
$\sigma = 1.4\, \mu\mathrm{m}$ \cite{Roati2017} (corresponding to $(\hbar/m c
\sigma)^2 \simeq 0.16$ at unitarity). Then the distance between the two peaks
$\Delta x \simeq 3.6\, \mu\mathrm{m}$ is larger than the spatial resolution of
current experiments \cite{bloch2012quantum,burchianti2015all}. The minimal value
$\gamma_\mathrm{min}$ that can be detected using dispersive waves is then a
priori determined by the maximal propagation time in a condensate of size $L$,
$t_\mathrm{max} = L/c$. Imposing that the maximal separation time is
$t_\mathrm{sep} = t_\mathrm{max}$, taking\footnote{Although box potentials have
not reached this size yet \cite{Zwierlein2017} the wave packet can bounce back
at the edges, resulting in a longer propagation length than the size of the
box.} $L= 250 \mu\mathrm{m}$, and $c \simeq 20\, \mu\mathrm{m/ms}$ at unitarity
\cite{joseph2007measurement}, we obtain
\begin{equation}
|\gamma_\mathrm{min}| \simeq 33 \frac{\hbar}{m c L} \underset{1/\kF 
|a|=0}{\simeq} 0.074, 
\end{equation}
smaller than the value predicted by the RPA at unitarity. To illustrate, the
time used in Fig.~\ref{fig:WaveUni} is $t_\mathrm{sep} \simeq \SI{20}{ms}$,
comparable to $t_\mathrm{max} = \SI{13}{ms}$. It should thus be possible to
determine the sign of $\gamma$ at unitarity within present capabilities. 

In the BCS regime $|\gamma|$ is much larger, such that shorter times are enough
to distinguish dispersive waves. For an initial width $\sigma \simeq 1.2\,
\mu\mathrm{m}$ ($(\hbar/m c \sigma)^2 \simeq 0.1$), as used in
Fig.~\ref{fig:WaveBCS}, the separation time is $t_\mathrm{sep} \simeq
\SI{0.22}{ms}$ according to the RPA.

\subsection{Staying in the linear regime}

Finally, we introduce a simple criterion on the perturbation amplitude $\zeta$
that guarantees that the waves do not enter the nonlinear regime described in
Sec.~\ref{sec:shockwaves}. Even if nonlinearity does not completely remove
dispersive effects, it probably forbids a precise measurement of $\gamma$. We
consider the nonlinear deformation of a wave packet to be significant when the
propagation time exceeds \cite{salasnich2011supersonic}
\begin{equation}
t_{\mathrm{nl}} = \frac{\sigma}{|c(\zeta) - c|}.
\end{equation}
Here, $c$ is the phase velocity of the low-momentum waves at density $\rho_0$
and $c(\zeta)$ the same velocity at density\footnote{At unitarity and in the BCS
limit, where $\mu$ is proportional to the Fermi energy $\epsilon_\mathrm{F}$,
one obtains from $m c^2 = \rho \dd \mu/\dd \rho$ that $c \propto \rho^{1/3}$ so
that $c(\zeta) = c (1+\zeta)^{2/3}$. In the BEC limit where $\mu \propto \rho$,
we have $c(\zeta) = c (1+\zeta)$.} $\rho=\rho_0(1+\zeta)^2$. When
$t=t_{\mathrm{nl}}$, the waves in the peak of perturbation have traveled a
distance $\sigma$ away from the wave packet, therefore causing deformations of
the wave front. In order for dispersive effects to be visible without nonlinear
deformations, the initial perturbation $\zeta$ should be sufficiently small so
that $t_{\mathrm{nl}}>t_{\mathrm{sep}}$. 

\begin{figure}
	\centering
	\includegraphics[scale=1]{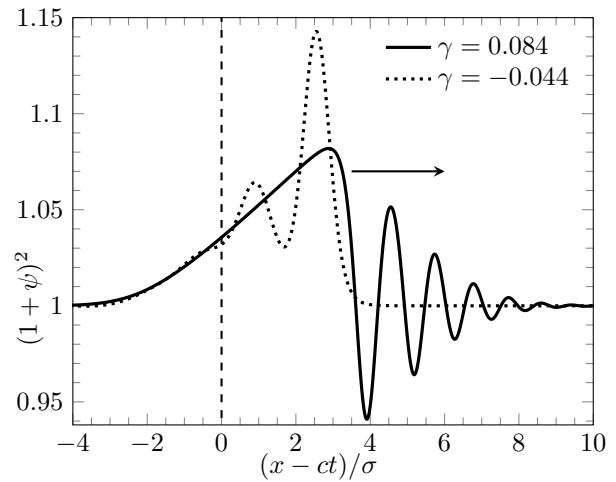}
	\caption{ \label{fig:UniNonLin} Wave propagation at unitarity according to
	Eq.~\eqref{eq:EFT}, where the coefficients are altered to incorporate two
	different predictions of $\gamma$ (solid line for the RPA prediction, dotted
	line for the prediction of Ref.~\cite{zou2017low}). Only the
	right-travelling wave is shown at $t = 0.5 t_\mathrm{sep} = 2.8
	t_\mathrm{nl}$, from an initial perturbation with $\zeta=0.1$ and $\sigma =
	2.5 \hbar/mc$.}
\end{figure}

Away from unitarity, where $|\gamma|$ is not anomalously small, this condition
is well satisfied for perturbations of order $\zeta \approx 5\%$, for which the
secondary peak $A_2$ has an amplitude of about 1\% of the background. A density
fluctuation of this magnitude is within experimental detectability
\cite{sidorenkov2013second,bloch2012quantum}. At unitarity, most theories
predict $|\gamma|$ to be very small, which results in a long dispersive
separation time. Therefore, fulfilling the condition
$t_{\mathrm{nl}}>t_{\mathrm{sep}}$ is possible only for very small
perturbations, experimentally challenging to prepare and observe. Still, the
sign of $\gamma$ can be assessed by entering the nonlinear regime with more
easily detectable larger perturbations. To clarify this, we extend the nonlinear
wave equation \eqref{eq:EFT} in order to describe a concave dispersion. The
coefficient $R$ is now chosen to reproduce the desired value of $\gamma$. In
Fig.~\ref{fig:UniNonLin}, we use this model to compare subsonic and supersonic
waves (with values of $\gamma$ as in Fig.~\ref{fig:WaveUni}) for an increase of
the superfluid density ($\zeta = 0.1$) sufficiently large to reveal nonlinear
effects. As with more usual nonlinear dispersive wave equations
\cite{el2016dispersive}, we observe that the orientation of the dispersive shock
wave (the position of the oscillatory train with respect to the main peak)
depends only on the sign of $\gamma$. This indicates that our scenario to
measure the sign of $\gamma$ is robust against nonlinear effects.

\section{Conclusion}

We have demonstrated that dispersive waves can be used as an alternative to
Bragg spectroscopy \cite{steinhauer2002excitation,hoinka2017goldstone} to
measure the first dispersive correction to the collective branch of a superfluid
Fermi gas. After a propagation time that we properly define, there appear
deformations either behind or in front of the wave front, depending on whether
the branch is subsonic or supersonic.

We show that using state-of-the-art experimental techniques it should be
possible to assess the nature of the dispersion at unitarity. Our study takes
into account possible nonlinear deformations and quantifies their relevance in
experimental conditions, which is rarely done for Fermi gases.

\begin{acknowledgments}
	
	Discussions with N.~Verhelst are gratefully acknowledged. W.V.A.
	acknowledges financial support in the form of a PhD fellowship of the Fonds
	Wetenschappelijk Onderzoek Vlaanderen (FWO). This research was supported by
	the Bijzonder Onderzoeksfonds by the Research Council of Antwerp University,
	the FWO project G.0429.15.N, and the European Union's Horizon 2020 research
	and innovation program under the Marie Sk\l{}odowska-Curie grant agreement
	number 665501.
	
\end{acknowledgments}

\appendix

\section{Equation of motion \label{AppEOM}}

The equation of motion \eqref{eq:WaveEq} can be rigorously derived in the
functional integral formalism. Expanding the full action of the system up to
quadratic order in phase $\theta$ and amplitude $\lambda$ fluctuations of the
order parameter gives the Gaussian fluctuation action 
\cite{diener2008quantum}
\begin{equation}
	\mathcal{S} = \mathcal{S}_0 + \int \dd \omega \sum\limits_{\bf{q}}  
		\begin{pmatrix}
			\lambda^\ast & \theta^\ast 
		\end{pmatrix} {\bf{M}}(\omega, \bf{q})
		\begin{pmatrix}
			\lambda \\ \theta 
		\end{pmatrix},
\label{eq:Squad}
\end{equation}
with ${\bf{M}}(\omega, \bf{q})$ the $2 \times 2$ Gaussian fluctuation matrix,
see Eqs. (38-39) in Ref.~\cite{Kurkjian2017}. From the action \eqref{eq:Squad}
coupled linear equations of motion can be derived for the phase and amplitude
fields. Alternatively, one can integrate out the amplitude field $\lambda$ in
the partition function
\begin{equation}
	\mathcal{Z} = \int \, \mathcal{D}\lambda \, \mathcal{D}\theta \, 
	\mathrm{e}^{-\mathcal{S}},
\end{equation}
which yields an effective action for the phase field $\theta$
\begin{equation}
	\tilde{\mathcal{S}} = \mathcal{S}_0 + \int \dd \omega \sum\limits_{\bf{q}} 
		\frac{\det {\bf{M}}(\omega, \bf{q})}{M_{1,1}(\omega, 
		\bf{q})}
			\theta^\ast \theta.
\end{equation}
As the zeros of $\det {\bf{M}}(\omega, \bf{q})$ describe the collective mode
dispersion $\omega_{\bf{q}}$, this action can always be written as
\begin{equation}
	\tilde{\mathcal{S}} = \mathcal{S}_0 + \int \dd \omega \sum\limits_{\bf{q}} 
		P(\omega,{\bf{q}}) (\omega^2 - \omega_{\bf{q}}^2) \theta^\ast 
		\theta,
\end{equation}
where $P(\omega,\bf{q})$ is some polynomial in $\omega$ and $\bf{q}$ that does
not vanish below the pair-breaking continuum. Extremizing the action ($\delta
\mathcal{S}/\delta \theta^\ast=0$), switching to the time domain, and
identifying $\psi$ to $\theta$ thus leads to Eq.~\eqref{eq:WaveEq} in the main
text.

\section{Effective field theory}

\label{app:EFT}
At zero temperature and in the 3D thermodynamic limit, the coefficients $C$,
$E$, $Q$, and $R$ of Eq.\eqref{eq:EFT} are given by \cite{Klimin2015}
\begin{align}
C &= \int \frac{\dd \bf{k}}{(2 \pi)^3} \frac{\hbar^4 k^2}{6 m^2} \frac{1}{4 \,  
\mathcal{E}_{\bf{k}}^3} \\
E &= \int \frac{\dd \bf{k}}{(2 \pi)^3} \frac{\hbar^4 k^2}{3 m^2} \frac{5 \, 
\xi_{\bf{k}}^2}{16 \, \mathcal{E}_{\bf{k}}^7} \\
Q &= \int \frac{\dd \bf{k}}{(2 \pi)^3} \frac{\hbar^2}{8 \, 
\mathcal{E}_{\bf{k}}^3} \\
R &= \int \frac{\dd \bf{k}}{(2 \pi)^3} \frac{ \hbar^2}{16 \, 
\mathcal{E}_{\bf{k}}^5} 
\end{align}
with $\vert \Delta \vert$ the bulk value of the superfluid order parameter and
$\xi_{\bf{k}} =  \frac{\hbar^2 k^2}{2 m} - \mu$ and $\mathcal{E}_{\bf{k}} =
\sqrt{\xi_{\bf{k}}^2 + \vert \Delta \vert^2}$ respectively the dispersion
relations of free fermions and of BCS quasiparticles. The functions $A$ and $D$
of the perturbed order parameter $\Psi$ are
\begin{align}
A(\vert \Psi \vert^2) &= - \frac{m}{4 \pi \hbar^2 a} - \frac{1}{2}\int 
\frac{\dd 
\bf{k}}{(2 \pi)^3}  \left( \frac{1}{E_{\bf{k}}} - \frac{2m}{\hbar^2 k^2} 
\right) \\
D(\vert \Psi \vert^2) &= \int \frac{\dd \bf{k}}{(2 \pi)^3}  \frac{\hbar \, 
\xi_{\bf{k}}}{4  E_{\bf{k}}^3} 
\end{align}
with $E_{\bf{k}} =\sqrt{\xi_{\bf{k}}^2 + \vert \Psi \vert^2}$.
\bibliography{ShockBib}

\end{document}